\documentstyle[twocolumn,prc,aps,psfig]{revtex}

     \newcommand{\pathnow}{}
\begin{document}\hbadness=10000
\twocolumn[\hsize\textwidth\columnwidth\hsize\csname %
@twocolumnfalse\endcsname
\title{Sudden Hadronization in Relativistic Nuclear Collisions}
\author{Johann Rafelski$^{a}$ and Jean Letessier$^{b}$}
\address{
Department of Physics$^{a}$, University of Arizona, Tucson, AZ 85721\\
and\\
Laboratoire de Physique Th\'eorique et Hautes Energies$^{b}$\\
Universit\'e Paris 7, 2 place Jussieu, F--75251 Cedex 05.
}
\date{June 19, September 18, 2000}
\maketitle
\begin{abstract}
\noindent
We formulate and study a mechanical instability criterion for sudden hadronization 
of dense matter fireballs formed in  158$A$ GeV  Pb--Pb collisions. 
Considering properties of quark-gluon matter and hadron gas  we obtain
the phase boundary between these two phases and 
demonstrate that the required deep QGP supercooling prior to sudden hadronization
has occurred. 

\vskip 0.2cm
PACS: 12.38.Mh, 25.75.-q, 12.40.Ee, 
\end{abstract}
\pacs{PACS: 12.38.Mh, 25.75.-q, 24.10.Pa}
\vspace{-0.35cm}
]
\begin{narrowtext}
Hot and dense hadron matter fireball is formed  in central 
collisions of relativistic 158$A$GeV heavy Pb-nuclei with Pb-target, 
comprising a new state of matter \cite{CERN}. Driven
by internal pressure, a  fireball expands and 
ultimately  a breakup (hadronization) into final state  particles occurs.
Early on in  the hadron production data analysis it was discovered that 
strange hadrons emerge from a source in which  strange $s$ and 
antistrange $\bar s$ quarks have same size phase space \cite{Raf91}. 
This is the case if final state  hadrons are directly produced by
the deconfined quark matter phase. 

The required
sudden fireball breakup could arise if a fireball made of a
the new form of matter significantly supercools, and in this 
state encounters a strong mechanical instability \cite{Cso94}. Despite
extensive ensuing study, the mechanisms determining when and
how sudden  hadronic particle abundances are formed
(chemical freeze-out) have not been fully understood \cite{Cse95,Bir99,Mis99}. 
However, there is growing evidence that 
the fireball breakup occurs 
over a relatively short period of time \cite{Ste99,Zab98,Sae97}. 

We propose and study here a natural mechanical instability criterion 
ensuing the fireball expansion into the metastable supercooled state.
We consider the exploding fireball dynamics in its center of momentum frame of reference.
The surface normal vector of exploding fireball is $\vec n$, and the local velocity 
of matter flow $\vec v_{\mbox{\scriptsize c}}$. 
The rate of  momentum flow vector $\vec {\cal P}$ 
at the surface  is obtained from the energy-stress tensor $T_{kl}$ \cite{LanHyd}: 
\begin{equation}\label{Peqv}
\vec {\cal P}=P^{(i)}\vec n+(P^{(i)}+\varepsilon^{(i)})
  \frac{\vec v_{\mbox{\scriptsize c}}\, \vec v_{\mbox{\scriptsize c}}\!\cdot\! \vec n}
          {1-\vec v_{\mbox{\scriptsize c}}^{\,2}}\,.
\end{equation}
The upper index  $(i)$ refers for the intrinsic energy density $\varepsilon$ and
pressure $P$  of matter in the  frame of reference,  
locally at rest, {\it i.e.} observed by a co-moving observer.  
We  omit the superscript  $(i)$ in the following.
 For the fireball expansion to continue, ${\cal P}\equiv |\vec {\cal P} |> 0$
is required.  For ${\cal P}\to 0$ at $v_{\mbox{\scriptsize c}}\ne 0$, we have
a conflict between the desire of the motion to stop or even reverse, and 
the continued inertial expansion. 

When the flow velocity remains 
large but ${\cal P}\to 0$, the intrinsic 
pressure $P$ must be negative. As 
illustration consider the fireball to be made of 
a quark-gluon liquid confined by an external vacuum 
pressure $\cal B$.  The total pressure 
and energy  comprise particle (subscript $\mbox{\scriptsize p}$)
and the  vacuum properties:
\begin{equation}\label{EPB1}
P=P_{\mbox{\scriptsize p}}-{\cal B}\,,\quad \varepsilon
 =\varepsilon_{\mbox{\scriptsize p}}+{\cal B}\,.
\end{equation}
Eq.\,(\ref{Peqv}) with $\vec {\cal P}=0$ thus reads:
\begin{equation}\label{BPeqv}
{\cal B}\vec n=P_{\mbox{\scriptsize p}}\vec n+
      (P_{\mbox{\scriptsize p}}+\varepsilon_{\mbox{\scriptsize p}})
\frac{\vec v_{\mbox{\scriptsize c}} \, \vec v_{\mbox{\scriptsize c}}\!\cdot \!\vec n}
        {1-v_{\mbox{\scriptsize c}}^{2}}\,,
\end{equation}
and it describes the (equilibrium) condition where the pressure of the expanding 
quark-gluon  fluid is just balanced by the external vacuum pressure.

Expansion beyond  ${\cal P}\to 0$ is in general not possible. 
A surface region of the fireball that  reached it but continues to flow outwards 
must   be torn apart. This is a collective instability and thus the 
ensuing disintegration of the fireball matter will be 
very rapid, provided that much of the surface reaches this condition.
We adopt the condition $\vec {\cal P}=0$ at any surface region to be
the  instability condition of an expanding hadron 
matter fireball.

Negative internal pressure $P<0$ is a requirement. 
At this stage the fireball must thus
be significantly supercooled.
The adiabatic transfer of internal heat  into accelerating 
flow of matter provides the mechanism which 
leads on the scale of $\tau=2\,10^{-23}$\,s to the development of this `deep'
supercooling 

It is possible to determine experimentally if
the  condition $P<0$ has been reached. Namely, the
 Gibbs-Duham relation for a unit volume:
\begin{equation}\label{1stlaw}
P=T\sigma+\mu_{\mbox{\scriptsize b}}\nu_{\mbox{\scriptsize b}}- \varepsilon\,,
\end{equation}
relates   the pressure, to  entropy 
density $\sigma=S/V$\,, energy density $\varepsilon=E/V$\,, 
and baryon density $\nu_{\mbox{\scriptsize b}}=b/V$\,, $V$ is the volume,
$T$ is the temperature, and $\mu_{\mbox{\scriptsize b}}$ the baryochemical potential.
Dividing by $\varepsilon$ we obtain:
\begin{equation}\label{EBStest}
\frac{PV}{E}=\frac{T_{\mbox{\scriptsize h}}}{E/S}
    +\frac{\mu_{\mbox{\scriptsize b}}}{E/b}-1\,.
\end{equation}
The microscopic processes governing the fireball
breakup determine how the quantities entering the right hand side of 
Eq.\,(\ref{EBStest}) are changed as hadrons emerge. Understanding this
we can determine, if the intrinsic fireball pressure  prior to  breakup,  
has been negative. 

The energy $E$  and baryon content  $b$ of the fireball are  conserved. 
Entropy $S$ is conserved when the gluon content of a QGP fireball is transformed into 
quark pairs in the entropy conserving  process $G+G\to q+\bar q$. Similarly, when
quarks and antiquarks recombine into hadrons,  entropy is conserved in the 
range of parameters of interest here.  Thus  also $E/b$ and $S/b$ is conserved 
across hadronization condition. The sudden hadronization process 
also maintains the temperature $T$ and baryo-chemical potential $\mu_b$
across the phase boundary. What changes are the chemical occupancy parameters. 
As gluons convert into quark pairs and hadrons $\gamma_g\to 0$ but
the number occupancy of light valance quark  pairs 
increases $\gamma_q>\gamma_{q_0}\simeq 1$
 increases significantly, along with  the number occupancy 
of strange quark  pairs $\gamma_s>\gamma_{s_0}\simeq 1$.

The sudden hadronization picture differs
from, {\it e.g.,} the  droplet-driven reequilibration 
transformation \cite{CK92,Dum97}, in which chemical equilibrium of
valance quark pair abundances is  maintained. 
Instead a change (reheating) of statistical
parameters $T,\mu_b$  occurs, along with a possible formation of a mixed phase 
required for volume expansion. In such reequilibration hadronization picture, 
entropy increase can also occur \cite{CK92b}. 
To draw the line  between these two hadronization pictures (equilibrium/sudden),
we need to determine the quark pair occupancy parameters $\gamma_i$. 
Our study of final state hadron abundances strongly favors 
$\gamma_q\simeq \exp (m_\pi/2T)>1, \gamma_s\simeq \gamma_i$  \cite{Let00}. 
Evaluating Eq.\,(\ref{EBStest}) using the results of our data analysis, 
we indeed obtain $P_f<0$. The magnitude of $|P_f|$  can vary between 
a few percent (in terms of energy density $E_f/V$), up to 20\% for the latest 
published result \cite{Let00}. The
precise value, which arises from several cancellations of larger numbers 
is sensitive to the strategy of how the currently available experimental 
data is described, {\it e.g.,} if strangeness  conservation is implemented, and 
if so, if  differentially at
each rapidity, or as an overall conservation law; how many high mass resonances 
can be excited in hadronization process, etc ... 

Importantly, we have not been able to obtain a scheme of hadron 
production analysis which describes the data with `$\chi^2/\mbox{dof}<1$
and would  not imply $P<0$ for the hadronizing fireball matter. On the
other hand, if we do force the hadronizing particles to 
be in chemical equilibrium, we find $\chi^2/\mbox{dof}>2.5,\ \mbox{dof}=10$ 
in our analysis which agrees for this limit with  \cite{BM99},
and in this case we find $P>0$.

Understanding in detail the breakup  condition ${\cal P}\to 0$ requires
that we model the shape and direction of flow in the late
stage of fireball evolution, obviously not an easy task. However,
considering $\vec n \!\cdot\! \vec {\cal P}\to 0$, we  find  
the constraint:
\begin{equation}\label{Pn}
\frac{-PV/E}{1+PV/E}=\kappa \frac{v_{\mbox{\scriptsize c}}^2}
               {1-v_{\mbox{\scriptsize c}}^2}\,, \quad 
  \kappa=(\vec v_{\mbox{\scriptsize c}}\! \cdot\! \vec n)^2/v_{\mbox{\scriptsize c}}^2\,.
\end{equation}
For an exactly symmetrical, spherical expansion the two vectors
$\vec v_{\mbox{\scriptsize c}}$ and $\vec n$  are everywhere parallel, thus $\kappa\to 1$.
However, in 158$A$ GeV Pb--Pb reactions  the longitudinal flow is 
considerably greater than the transverse flow \cite{NA49}, and we note
$\kappa\to 0$ for a longitudinally evolving cylindrical 
fireball. For the Pb--Pb collisions considered here, our analysis 
suggest $0.1<\kappa<0.6$.

We now substitute, in Eq.\,(\ref{Pn}), the fireball matter properties 
employing the  Gibbs-Duham relation, Eq.\,(\ref{1stlaw}), and arrive at: 
\begin{equation}\label{EBSfinal}
\frac{E}{S}=\left(T_{\mbox{\scriptsize h}}+\frac{\mu_{\mbox{\scriptsize b}}}{S/b}\right)
     \left\{1+\kappa \frac{v_{\mbox{\scriptsize c}}^2}
                 {1-v_{\mbox{\scriptsize c}}^2}\right\}\,.
\end{equation}
Eq.\,(\ref{EBSfinal}) establishes a general 
constraint characterizing the fireball breakup condition. 

The solid line, in figure~\ref{vtfig}, shows  the 
behavior of $v_{\mbox{\scriptsize c}}(T_{\mbox{\scriptsize h}})$ 
constraint arising from Eq.\,(\ref{EBSfinal})
for the example $E/S=0.184\pm0.05$\,GeV (error range shown 
by dotted lines), $\kappa=0.6$. Outside  of the  region bounded by the solid line 
({\it i.e.,} for greater $T_{\mbox{\scriptsize h}}$ 
and $v_{\mbox{\scriptsize c}}$), the flow expansion can occur as the
internal particle pressure is greater than the confining pressure. 
Also shown  in figure~\ref{vtfig} is hadron production analysis result\cite{Let00} and
its statistical error, the systematic error is of same magnitude.
The agreement of theory and experiment  results from the choice of non-spherical flow 
with specific freeze-out  shape described by the average value $\kappa=0.6$,
see Eq.\,(\ref{Pn}). Figure \ref{vtfig} illustrates the 
great sensitivity to the  analysis on the freeze-out constraint.
The dashed  horizontal 
line, in  figure \ref{vtfig}, is the velocity of sound of the interacting 
quark-gluon liquid, which barely differs  from $1/\sqrt{3}$ \cite{Ham00}. 

\begin{figure}[tb]
\vspace*{-3.3cm}
\hspace*{-1.cm}\psfig{width=11cm,clip=,figure=\pathnow 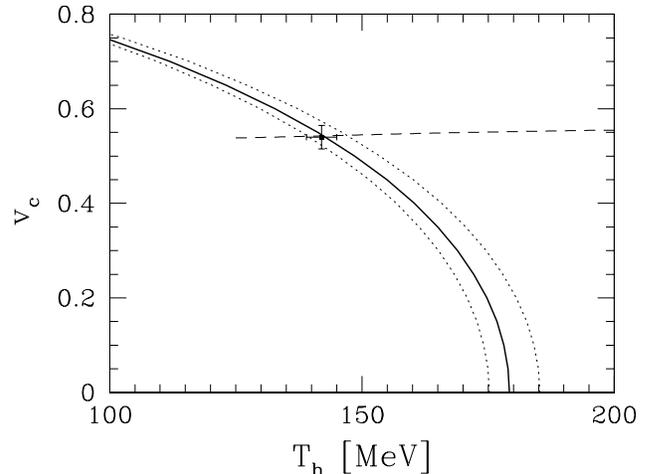}
\vspace*{-1.9cm}
\caption{ 
Fireball velocity as function of breakup temperature constraint for
 the case $E/S=0.185\pm0.005$\,GeV, with $S/b=42$ and $\mu_{\mbox{\scriptsize b}}=0.2$\,GeV; 
the dotted lines
describe the uncertainty in the determination of $E/S$. 
Dashed line: velocity of sound of relativistic quark-gluon liquid. Also shown is
hadron production analysis result \protect\cite{Let00}.
\label{vtfig}
}
\end{figure}

We have so far not used in the discussion any key specific property of 
the equations of state of the matter filling the fireball.
However, our results imply that the matter inside the fireball is 
deeply supercooled. Can this be the deeply supercooled 
liquid of quarks  and gluons? In fact a study of
QGP equations of state  employing properties of 
QCD interactions and thermal QCD \cite{Ham00}, fine tuned to
agree with the properties of lattice QCD results \cite{Kar00}
suggest that. We extend this  study to consider the phase boundary. 
The thin solid line in the $T,\mu_{\mbox{\scriptsize b}}$ plane  
in figure \ref{PLMUPLIQ} shows  where the pressure of the quark-gluon
liquid  equals the equilibrated hadron gas pressure. 
The hadron gas behavior is obtained evaluating and summing 
the contributions  of  all known hadronic resonances considered to be
point particles. When we allow for 
finite volume of hadrons \cite{HR80}, we find that the hadron  pressure is
slightly reduced, leading to some (5\,MeV) reduction in the  equilibrium transition 
temperature, as is shown by  the dashed line in figure \ref{PLMUPLIQ}\,.
For vanishing baryo-chemical potential, 
we note in figure \ref{PLMUPLIQ} that the equilibrium
phase transition temperature is  $T_{\mbox{\scriptsize pt}}\simeq 172$\,MeV,
and when finite hadron size is allowed, $T_{\mbox{\scriptsize fp}}\simeq 166$\,MeV,
The scale in temperature we discuss is result of 
comparison with lattice gauge results. 
Within the lattice calculations \cite{Kar00}, 
it arises from the comparison with  the string tension.

\begin{figure}[tb]
\vspace*{-2cm}
\hspace*{-1cm}\psfig{width=10.5cm,clip=,figure=\pathnow 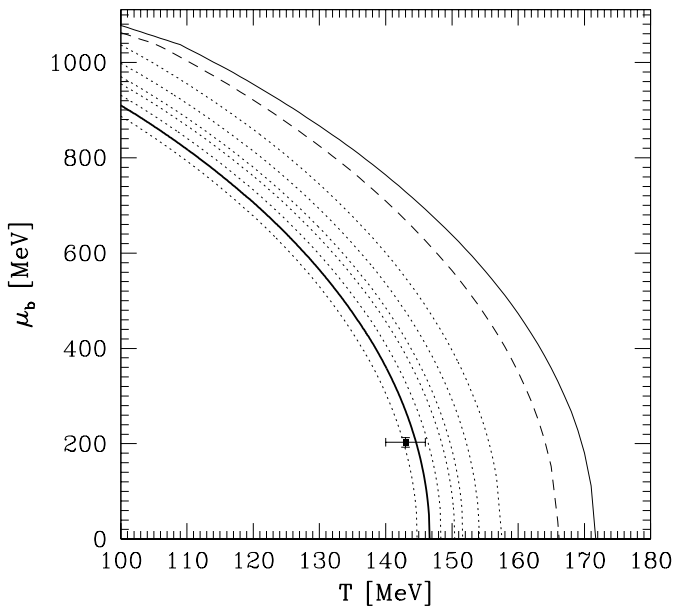}
\vspace*{-0.8cm}
\caption{ 
Thin solid and dashed lines: equilibrium phase transition from
hadron gas to QGP liquid without and with excluded volume correction,
respectively.  Dotted:  
breakup condition at shape parameter $\kappa=0.6$, for expansion velocity 
$v_{\mbox{\scriptsize c}}^2=0, 1/10, 1/6, 1/5, 1/4$ and 1/3, and thick line for 
$v_{\mbox{\scriptsize c}}=0.54$. The experimental point denotes 
chemical nonequilibrium freeze-out analysis result~\protect\cite{Let00}.
\label{PLMUPLIQ}
}
\end{figure}

The dotted lines, in figure \ref{PLMUPLIQ}, correspond to 
the condition Eq,\,(\ref{BPeqv}) using the shape parameter 
$\kappa=0.6$, Eq.\,(\ref{Pn}),  for (from right to left) 
$v_{\mbox{\scriptsize c}}^2=0, 1/10, 1/6, 1/5, 1/4$ and $1/3$. The
last dotted line corresponds thus to an expansion flow with 
the velocity of sound of  relativistic noninteracting massless gas. The thick
solid line corresponds to an expansion with $v_{\mbox{\scriptsize c}}=0.54$.  
The hadron analysis result is also shown \cite{Let00}. 
Comparing in figure \ref{PLMUPLIQ} 
thin solid/dashed with the  thick line,
we recognize the deep supercooling as required for the explosive 
fireball disintegration. The super-cooled  zero pressure $ P=0$ QGP  
temperature is at $T_{\mbox{\scriptsize sc}}=157$\,MeV, 
(see the intercept of the first dashed line to the right 
in figure \ref{PLMUPLIQ}) and an expanding fireball can deeply 
super-cool to $T_{\mbox{\scriptsize dsc}}\simeq 147$\,MeV 
(see the intercept of thick solid line)
before the mechanical instability occurs.

Deep supercooling requires a  first order phase transition, and
this in turn implies presence of latent heat $\cal B$. 
Physical consistency then requires  presence of 
external (negative) vacuum pressure $-{\cal B}$. 
More precisely, the vacuum contribution to the physical properties 
of deconfined matter can be derived from
$\ln{\cal Z}_{\mbox{\scriptsize vac}}\equiv -{\cal B}V\beta$\,:
\begin{eqnarray}\label{PVac}
P_{\mbox{\scriptsize vac}}&=&\frac{T}{V}\ln{\cal Z}_{\mbox{\scriptsize vac}}
     =-{\cal B}\,,\\ \label{EVac}
\varepsilon_{\mbox{\scriptsize vac}}&=&
   -\frac{\partial\ln {\cal Z}_{\mbox{\scriptsize vac}}}{V\partial \beta}=
{\cal B}\left\{1+\frac{\partial \ln {\cal B}/{\cal B}_0}
      {\partial \ln \beta/\beta_0}\right\}\,.
\end{eqnarray}
The temperature, $T=1/\beta$, dependence of the vacuum pressure
 has been considered within the model of color-magnetic vacuum 
structure \cite{Mul81,Kap81}. Near to the phase 
transformation condition, the variation of $\cal B$ with $\beta$ 
is minimal (see figure 2 in \cite{Mul81}),
and thus  the logarithmically small last term in Eq.\,(\ref{EVac})
can be, in principle, ignored.

We now combine the theoretical properties of the QGP equations of state 
with the dynamical fireball properties in order to constrain ${\cal B}$. 
Reviewing Eq.\,(\ref{EBStest}), we obtain:
\begin{equation}\label{EBSres}\label{EBSineq}
-\frac{PV}{E}\varepsilon_{\mbox{\scriptsize QGP}}
+P_{\mbox{\scriptsize p}}={\cal B}\,,
\end{equation}
To evaluate $\cal B$, we note that lattice results for $\varepsilon_{\mbox{\scriptsize QGP}}$ 
are well represented by $\varepsilon_{\mbox{\scriptsize QGP}}=aT^4$,  with $a\simeq 11$, 
value extrapolated for the number of light quark flavors being $n_f=2.5$ at the 
hadronization point \cite{Ham00}. 
We obtain, for the fireball formed in Pb--Pb reactions,
$$0.2 \cdot  {11} T_{\mbox{\scriptsize h}}^4\simeq 0.17\mbox{ GeV/fm}^3\le {\cal B}\,.$$

Is our picture of fireball evolution compelling?
We found that particle production  occurred at condition of 
negative  pressure expected in a deeply 
supercooled state and have shown 
internal  consistency with (strange) hadron  production
analysis involving chemical non-equilibrium. Moreover, 
these chemical freeze-out conditions agree with
thermal analysis \cite{Ste99}, allowing the conjecture that the 
explosive quark-gluon fireball breakup forms final state hadrons, which do not 
undergo further reequilibration. However, we noted that the chemical equilibrium 
reaction picture differs from ours only in terms of its
statistical significance ($\chi^2/\mbox{dof}>2.5$, \cite{BM99}). It produces
a chemical freeze-out temperature of $T=168$\,MeV, just the value we found for
an equilibrium phase transition implicit in the assumption of chemical equilibrium.
The higher chemical freeze-out temperature produces greater population of excited
hadronic states. Their  decays deform hadron spectra, and this allows for a second 
evolution stage with a thermal freeze-out  temperature  at or below $120$\,MeV. 

Strongly in favor of the here described sudden QGP hadronization resulting in chemical
nonequilibrium reaction picture is the presence of a  hadron 
multiplicity excess, related to an entropy excess
 \cite{Let93}.  This is seen both as multiplicity per baryon, and as 
increase of multiplicity comparing $pA$ to $AA$ reactions. 
We could not describe this  effect by 
admitting other physical models  such as, in medium, change of hadron masses.
We have found that invariably the statistical significance of the analysis 
decreases as we modify individual hadron properties in an ad-hoc fashion, while
maintaining chemical equilibrium of hadron abundances. Our experience shows that 
the only theoretical description of
 hadron production data that  works ($\chi^2/\mbox{dof}<1$) requires excess of 
valence quark pair abundance, irrespective of the detailed strategy of data analysis.
This agrees well with the dynamical study of nuclear collisions within the UrQMD 
model  concludes that at CERN energies the chemical non-equilibrium is 
required to characterize the numerical results in terms of a statistical 
model \cite{Bra99}.

Also in favor of our result is the conclusion of  Cs\"org\H{o} and Csernai  \cite{Cso94}:
who required 
as verification for the presence of a deeply supercooled state of matter and sudden 
hadronization: 
i) short duration and relatively short mean proper-time of 
particle emission, now seen in particle correlations  \cite{Ste99},
ii) clean strangeness signal of QGP \cite{WA97p};
iii) universality of produced particle spectra
which are the remarkable features of strange particle production \cite{WA97s} 
v) no mass shift of the phi-meson; despite extensive search such a 
 shift has not been found by the NA49 collaboration \cite{NA49phi}.

In summary, we  have introduced a constraint,  Eq.\,(\ref{EBSfinal}), 
which relates the physical and statistical properties 
of the hadronic fireball at the point of sudden breakup. 
We obtained this result from mechanical stability consideration, 
employing only  properties of the energy-stress tensor of 
matter,  and the Gibbs-Duham relation, Eq.\,(\ref{1stlaw}). 
We showed that this constraint is consistent with analysis results 
obtained considering the experimental particle production data
for Pb--Pb collisions at 158$A$ GeV.
We further studied the behavior of the phase 
transition between hadron gas and quark-gluon liquid, and have determined the 
magnitude of the deep supercooling occurring in the fireball expansion.
Employing a lattice-QCD based estimate on number of
degrees of freedom in the energy density of the QCD thermal matter, we obtained 
a constraint on the magnitude of latent heat/vacuum pressure 
${\cal B}\ge 0.17\mbox{\,GeV/fm}^3$.  We conclude that both 
in theoretical study of the data as well as for reasons of principle the 
deciding factor about the sudden nature of the 
phase transformation is the absence of chemical equilibrium.

%
We thank T. Cs\"org\H{o} for constructive comments.
Work Supported in part by a grant from the U.S. Department of
Energy,  DE-FG03-95ER40937\,. LPTHE, Univ.\,Paris 6 et 7 is:
Unit\'e mixte de Recherche du CNRS, UMR7589.




\end{narrowtext}


\end{document}